\documentclass[12pt]{iopart}

\usepackage{iopams}
\usepackage{amsthm}
\usepackage{graphicx}

\usepackage{dsfont}

\theoremstyle{definition}

\theoremstyle{plain}

\theoremstyle{remark}

\newtheorem*{Thm*}{}

\newcommand{\dd}{\mathrm{d}}
\newcommand{\ee}{\mathrm{e}}
\newcommand{\ii}{\mathrm{i}}
\newcommand{\rh}{r_\mathrm{h}}
\newcommand{\Hp}{{\mathcal H_\mathrm{p}}}
\newcommand{\Hf}{{\mathcal H_\mathrm{f}}}
\newcommand{\Qp}{Q_\mathrm{p}}

\newcommand{\Al}{\mathcal A_1}
\newcommand{\Ar}{\mathcal A_2}
\newcommand{\E}{\mathcal E}
\newcommand{\Phibold}{{\bf\Phi}}
\newcommand{\p}{_\mathrm{p}}
\newcommand{\f}{_\mathrm{f}}

\newcommand{\R}{\mathds R}
\newcommand{\C}{\mathds C}

\newcommand{\N}{\mathds N}

\newcommand{\diff}[2]{\frac{\partial#1}{\partial#2}}

\newcommand{\beq}{\begin{equation}}
\newcommand{\eeq}{\end{equation}}
\newcommand{\bea}{\begin{eqnarray}}
\newcommand{\eea}{\end{eqnarray}}

\begin{document}

\title
{Regularity of Cauchy horizons in $S^2\times S^1$ Gowdy spacetimes} 

\author{J\"org Hennig$^1$ and Marcus Ansorg$^2$}
\address{$^1$ Max Planck Institute for Gravitational Physics,
Am M\"uhlenberg 1, D-14476 Golm, Germany}
\address{$^2$ Institute of Biomathematics and Biometry,
 Helmholtz Zentrum M\"unchen,
 Ingolst\"adter Landstr. 1,
 D-85764 Neuherberg, Germany}
\eads{\mailto{pjh@aei.mpg.de}, \mailto{marcus.ansorg@helmholtz-muenchen.de}}

\date{\today}

\begin{abstract}
We study general $S^2\times S^1$ Gowdy models with a regular past
Cauchy horizon  
and prove that a second (future) Cauchy horizon exists, provided that a
particular conserved 
quantity $J$ is not zero. We derive an explicit expression for the metric
form on the future Cauchy horizon in terms of the initial data on the
past horizon and conclude 
the universal relation $A\p A\f=(8\pi J)^2$ where $A\p$ and $A\f$ are
the areas of  
past and future Cauchy horizon respectively.

\end{abstract}



\section{Introduction \label{sec:Intro}}
The well-known singularity theorems by Hawking and Penrose \cite{Hawking}
show that
cosmological solutions to the Einstein equations generally contain
singularities.
%
%
 As discussed by Clarke \cite{Clarke} (see also \cite{Ellis} for a
comprehensive overview) there are two types of singularities:
(i) \emph{curvature singularities}, for which components of the Riemann
tensor or its $k$th derivatives are irregular (e.g. unbounded),
and (ii) \emph{quasiregular singularities}, which are associated with peculiarities
in the topology of space-time (e.g. the vertex of a cone), although the local geometry is
well behaved. In addition, the curvature singularities are divided up into \emph{scalar
singularities} (for which some curvature invariants are badly behaved)
and \emph{nonscalar singularities} (for which arbitrarily large or
irregular tidal forces occur).
The singularity theorems mentioned above provide, however, in general no information 
about the specific type of singularity --- they make statements solely
about causal geodesic incompleteness. This lack of knowledge 
concerning the specific nature of the singular structure
is the reason for many open outstanding problems in
general relativity, including the strong cosmic censorship conjecture
and the BKL conjecture (see \cite{Andersson} for an overview).

A major motivation for the study of \emph{Gowdy spacetimes} as
relatively simple, but non-trivial inhomogeneous cosmological models
results from the desire to understand the
mathematical and physical properties of such cosmological singularities.
The Gowdy cosmologies, first studied in \cite{Gowdy1971,Gowdy1974}, are characterized by an
Abelian isometry group $U(1)\times U(1)$ with spacelike group orbits, i.e.~these spacetimes
possess two associated spacelike and commuting Killing vector
fields
$\xi$ and $\eta$. Moreover, the definition of Gowdy spacetimes
includes that the \emph{twist constants}
$\epsilon_{\alpha\beta\gamma\delta}\xi^\alpha\eta^\beta\nabla^\gamma\xi^\delta$and
$\epsilon_{\alpha\beta\gamma\delta}\xi^\alpha\eta^\beta\nabla^\gamma\eta^\delta$
(which are constant as a consequence of the field equations) are
zero\footnote[1]{The assumption of vanishing twist constants is non-trivial
only in the case of spatial $T^3$ topology. Note that in spatial $S^3$ or $S^2\times S^1$ 
topology there are specific axes on which one of the Killing vectors vanishes 
identically, which leads to vanishing twist constants.}.

For compact, connected, orientable and smooth three manifolds, the corresponding spatial topology
must be either $T^3$, $S^3$,  $S^2\times S^1$ or $L(p,q)$, cf.~\cite{Gowdy1974} (see also 
\cite{Mostert,Neumann,Fischer}). Note that the universal cover of the lens space
$L(p,q)$ is $S^3$ and hence this case needs not be treated
separately, see references in \cite{Chrusciel1990}.

In the $T^3$-case, global existence in time with respect to the areal
foliation time $t$ 
was proved by Moncrief \cite{Moncrief1981}.
Moreover, he has shown that the trace of the second fundamental
form blows up uniformly on the hypersurfaces $t=\textrm{constant}$ in
the limit $t\to0$. As a consequence, the solutions do not permit a globally
hyperbolic extension beyond the time $t=0$. However, to date it has not 
been clarified whether the solutions are extendible (as
non-globally hyperbolic $C^2$-solutions) or are generically subject to curvature
singularities at $t=0$.

Although global existence of solutions inside the ``Gowdy square'' (i.e. for $0<t<\pi$, cf.~Fig.~\ref{GS} below) 
was shown by Chru\'sciel for $S^2\times S^1$ and $S^3$ topology, see Thm.~6.3 in \cite{Chrusciel1990}, 
it is still an open question whether globally hyperbolic extensions beyond the hypersurfaces $t=0$ or $t=\pi$ exist. 
It is expected that these hypersurfaces
contain either curvature singularities or Cauchy horizons; the
theorem in \cite{Chrusciel1990} however does not in fact exclude the
possibility that these are merely coordinate singularities.   

For \emph{polarized} Gowdy models, where the Killing vector fields
can be chosen to be orthogonal everywhere, the nature of the singularities
for all possible spatial topologies  has been studied in
\cite{Isenberg1990,ChruscielIM}.
In particular, strong cosmic censorship
and a version of the BKL conjecture have been proved.
Investigations of singularities in the \emph{unpolarized} case
for $T^3$ topology 
can be found in
\cite{Berger1993,Kichenassamy1998,Ringstrom2006,Ringstrom2006b}.
 
For unpolarized $S^3$ or $S^2\times S^1$ Gowdy spacetimes
not many results on singularities (strong cosmic
censorship, BKL conjecture, Gowdy spikes) are known.
Particular singular solutions have been constructed with Fuchsian
techniques in \cite{Stahl}. Moreover, numerical studies indicate
that the behavior near singularities and the appearance of spikes are similar 
to the $T^3$-case \cite{Garfinkle1999,Beyer2008,Beyer2009}.

In this paper, we study general (unpolarized or polarized)
$S^2\times S^1$ Gowdy models with a
regular Cauchy horizon (with $S^2\times S^1$ topology)
at $t=0$ (cf.~Fig.~\ref{GS})\footnote{Without loss of generality we choose a
\emph{past} Cauchy horizon $\Hp$.} 
and assume that the spacetime 
is regular (precise regularity requirements are given below)
at this horizon as well as in a neighborhood. As mentioned above, a theorem by
Chru\'sciel \cite{Chrusciel1990} implies then that the metric is regular
for all $t<\pi$, i.e.~excluding  
only the future hypersurface $t=\pi$. With the methods utilized 
in this paper we are able to provide the missing piece, 
i.e.~we prove that under our regularity assumptions the existence
of a regular second (future) Cauchy horizon $\Hf$ (at $t=\pi$) is implied, 
provided that a particular conserved
quantity $J$ is not zero\footnote{As we will see in
Sec.~\ref{Sec:conserved}, the conserved quantity
$J$ vanishes in \emph{polarized} Gowdy models.}.
Moreover, we derive an explicit 
expression for the metric form on the future Cauchy horizon in
terms of the initial data on the past horizon. From this explicit formula,
the universal relation $A\p A\f=(8\pi J)^2$
between the areas $A\p, A\f$ of past and future Cauchy horizons
and the above mentioned conserved quantity $J$ can be concluded.

The proofs of these statements can be found by relating any
$S^2\times S^1$  
Gowdy model to a corresponding axisymmetric and stationary black hole solution 
(with possibly non-pure vacuum exterior, e.g.~with surrounding matter), 
considered between outer event and inner Cauchy horizon. 
Note that the region between these horizons is regular hyperbolic,
i.e. the Einstein equations are hyperbolic PDEs in an appropriate gauge
with coordinates adapted to the Killing vectors, see
\cite{Ansorg2008,Ansorg2009,Hennig2009}\footnote{The interior of axisymmetric and stationary black hole solutions is
non-compact and has spatial
$S^2\times\R$ topology. Here the $\R$-factor is generated by
a subgroup of the symmetry group corresponding to one of the Killing
fields. Therefore, it is possible to factor out a discrete subgroup such
that $S^2\times S^1$ topology is achieved.}. 
(The Kerr metric is an explicitly known solution of these PDEs, see
\cite{Obregon}.)\footnote{
Another interesting example of a
spacetime with a region isometric to Kerr is the Chandrasekhar and
Xanthopoulos solution \cite{Chandrasekhar}
which describes colliding plane waves. It
turns out that  the region of interaction of the two waves is an alternative
interpretation of a part of the Kerr spacetime region between event
horizon and Cauchy horizon, cf. \cite{Griffiths,Helliwell}.}
As a consequence, the
results on the regularity of the interior of such black holes and
existence of regular Cauchy horizons inside the black holes
obtained in \cite{Ansorg2008,Ansorg2009,Hennig2009} can be carried over to
Gowdy spacetimes.

The results in \cite{Ansorg2008} were found by utilizing a particular soliton method --- the so-called
\emph{B\"acklund transformation}. Making use of	the theorem by
Chru\'sciel mentioned earlier, it was possible to show that 
a regular Cauchy horizon inside the black hole always exists, provided that the
angular momentum of the black hole does not vanish. (The above
quantity $J$ is the Gowdy counterpart of the angular
momentum.) 

In \cite{Ansorg2009,Hennig2009} these results have been generalized to the case in which an
additional Maxwell field is considered. The corresponding technique, that is the 
\emph{inverse scattering method}, again comes from soliton theory and permits the reconstruction of the field quantities 
along the entire boundary of the Gowdy square. Hereby, an associated linear matrix problem 
is analyzed, whose integrability conditions are equivalent to the non-linear field equations 
in axisymmetry and stationarity. Note that in this article we restrict ourselves to the pure 
Einstein case (without Maxwell field) and refer the reader to \cite{Ansorg2009,Hennig2009} 
for results valid in full Einstein-Maxwell theory.

We start by introducing appropriate coordinates, adapted to
the description of regular axes and Cauchy horizons at the boundaries of
the Gowdy square, see Sec.~\ref{Sec:coords}. Moreover, we revisit the
complex Ernst formulation of the field equations and corresponding 
boundary conditions and introduce the conserved
quantity $J$ in question. In this formulation we can translate
the results of \cite{Ansorg2008,Ansorg2009,Hennig2009} and obtain the metric 
on the future Cauchy horizon in terms of initial data on the past horizon, see Sec.~\ref{Sec:EP}.
As another consequence we arrive at the above equation relating $A\p,
A\f$ and $J$, see Sec.~\ref {Sec:formula}. 
Finally, in Sec.~\ref {Sec:Disc} we conclude with a discussion of our results.

\section{Coordinates and Einstein equations\label{Sec:coords}}
\subsection{Coordinate system, Einstein equations and regularity requirements\label{Subsec:coords}}

We introduce suitable coordinates and metric functions by adopting our
notation from \cite{Garfinkle1999}. Accordingly, we write the Gowdy line
element in the form
\begin{equation}\label{LE}
 \dd s^2=\ee^M(-\dd t^2+\dd\theta^2)+\sin t\sin\theta
         \left[\ee^L(\dd\varphi+Q\dd\delta)^2+\ee^{-L}\dd\delta^2\right],
\end{equation}
where the metric functions $M$, $L$ and $Q$ depend on $t$ and $\theta$
alone. In these coordinates, the two Killing vectors are given by
\begin{equation}
 \eta = \diff{}{\varphi},\qquad \xi = \diff{}{\delta}.
\end{equation}

As mentioned in Sec.~\ref{sec:Intro}, any $S^2\times S^1$ Gowdy model can be related to the spacetime portion
between outer event and inner Cauchy horizon of an appropriate axisymmetric and stationary black hole solution. 
Black hole spacetimes of this kind have been studied by Carter \cite{Carter} and Bardeen \cite{Bardeen}. Among 
other issues they discussed conditions for regular horizons. 
In this paper we adopt their regularity arguments for our study of Gowdy 
spacetimes. Accordingly we rewrite the line element (\ref{LE}) in the form 
\begin{equation}\label{LE2}
 \dd s^2 = \ee^M(-\dd t^2+\dd\theta^2)
           +\ee^u\sin^2\!\theta(\dd\varphi+Q\dd\delta)^2
           +\ee^{-u}\sin^2\! t\,\dd\delta^2          
\end{equation}
where
\begin{equation}\label{u}
 u=\ln\sin t-\ln\sin\theta + L.
\end{equation}
Now, at a {\em regular} horizon (clear statements about the type of regularity follow below) 
the metric functions $M,Q$ and $u$ are regular, meaning that $L$ possesses a specific irregular 
behavior there.\footnote{We achieve the form of the line element used in
\cite{Ansorg2008,Ansorg2009,Hennig2009} from \eref{LE2}
by introducing the Boyer-Lindquist-type coordinates
$(R,\theta,\varphi,\tilde t)$ with
$R:=\rh\cos t$, $\tilde t:=\delta/(2\rh)$, $\rh=\textrm{constant}$,
and the metric functions
$\hat\mu:=\ee^M$, $\hat u := \ee^u$, $\omega:=-2\rh Q$. Since the potentials
$\hat\mu>0$, $\hat u>0$ and $\omega$ are regular at the axes and at
the Cauchy horizon
(cf. \cite{Bardeen}), we see that $M$, $u$ and $Q$ are regular as well.} 

At this point, some remarks about the specific regularity requirements 
needed in our investigation are necessary. A crucial role is played by
a theorem of Chru\'sciel (Theorem 6.3 in \cite{Chrusciel1990})
which provides us 
with the essential regularity information valid in the {\em interior} 
of the Gowdy square. In this theorem it is assumed that initial data 
are given on an interior Cauchy slice, described by 
$t=\mbox{constant}=t_0,$ $0 < t_0 <\pi$. These data 
are supposed to consist of (i) metric potentials that are
$H^k$-functions of $\theta$
and (ii) first time derivatives that are $H^{k-1}$-functions of $\theta$
(with $k\ge3$). Here $H^k$ denotes the Sobolev space $W^{k,2}$ that contains all functions 
for which both the function and its weak
derivatives up to the order $k$ are in $L^2$.
With these assumptions the theorem by Chru\'sciel
guarantees the existence of a unique continuation of the given initial
data for which the metric is $H^k$ on 
all future spatial slices $t=\mbox{constant}$ with $t_0<t<\pi$,
i.e.~only the future boundary $t=\pi$ of the Gowdy square is excluded.
(Note that Theorem 6.3 as formulated in \cite{Chrusciel1990} assumes
the metric to be smooth. However, this condition can be
relaxed considerably to the assumption of $H^k$ spaces \cite{Chrusciel}.) 

\begin{figure}
 \centering
 \includegraphics[scale=0.7]{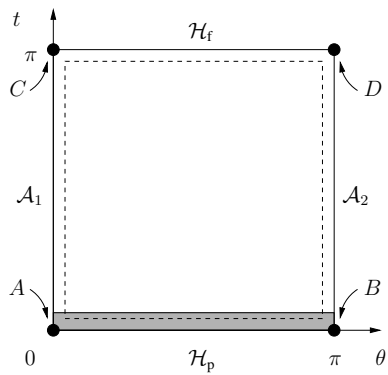}
 \caption{The Gowdy square. We assume a $H^4$-regular metric and
 $H^3$-regular time derivatives on all slices
 $t=\textrm{constant}$ in a neighborhood (gray region) 
 of the past Cauchy horizon ($\Hp: t=0$) and find
 by virtue of the results in \cite{Ansorg2008}, that then the metric 
 is $H^4$-regular on all future slices $t=\textrm{constant}$, $0\le t\le\pi$
 (unless the quantity $J$ introduced in (\ref{def_J}) is zero). In
 particular, a $H^4$-regular future Cauchy horizon ($\Hf: t=\pi$) exists.}
 \label{GS}
\end{figure}

Now, for the applicability of our soliton methods it is essential that the
metric potentials in \eref{LE2} possess $C^2$-regularity. 
Therefore, in order to apply both Chru\'sciel's theorem and the soliton
methods, we need to require that the metric potentials $M$, $u$, $Q$ be
$H^4$-functions and the time derivatives $H^3$-functions
of $\theta$ on all slices $t=\textrm{constant}$ 
in a neighborhood of the horizon $\Hp$, see Fig.~\ref{GS}.\footnote{In 
\cite{Ansorg2008,Ansorg2009,Hennig2009}, the much stronger
assumption was made that the metric functions be \emph{analytic}
in an {\em exterior} neighborhood of the black hole's event
horizon. This stronger requirement  was necessary to conclude that the
metric is also regular (in fact analytic) in an {\em interior}  
vicinity of the event horizon, a requirement needed for applying
Chru\'sciel's theorem.}  Then Chru\'sciel's theorem ensures 
the existence of an $H^4$-regular continuation which implies (via Sobolev
embeddings and the validity of the Einstein equations)
that the metric potentials $M$, $u$, $Q$ are $C^2$-functions of $t$ and $\theta$ for
($t,\theta)\in(0,\pi)\times[0,\pi]$, i.e.~in the entire Gowdy square
with the exception of 
the two horizons $\Hp\,(t=0)$ and $\Hf\,(t=\pi)$. Now, in accordance with Carter's and Bardeen's arguments 
concerning regularity at the horizon, we require 
that this $C^2$-regularity 
holds also for $t=0$, i.e.~we assume in this 
manner a specifically regular past horizon $\Hp$. 


As mentioned above, these requirements allow us to utilize our soliton methods at $\Hp$. 
Since $\Hp$ is a degenerate boundary surface of the interior hyperbolic region, 
the study of the Einstein equations provides us with specific relations that permit the identification of 
an appropriate set of initial data of the hyperbolic problem at the past Cauchy horizon $\Hp$.

For the line element \eref{LE2}, the Einstein equations read as follows:
\begin{equation}\fl\label{E1}
 -u_{,tt}-\cot t\, u_{,t} + u_{,\theta\theta}+\cot\theta\,u_{,\theta}
 = 2-\frac{\sin^2\theta}{\sin^2 t}\ee^{2u}
 \left(Q_{,t}^2 - Q_{,\theta}^2\right),
\end{equation}
\begin{equation}\fl\label{E2}
 -Q_{,tt}+\cot t\,Q_{,t}+Q_{,\theta\theta}+3\cot\theta\,Q_{,\theta}
 -2(u_{,t}Q_{,t}-u_{,\theta}Q_{,\theta})=0,
\end{equation}
\begin{equation}\fl\label{M}
 -M_{,tt}+M_{,\theta\theta}-\frac{1}{2}u_{,t}(u_{,t}-2\cot t)
 +\frac{1}{2}u_{,\theta}(u_{,\theta}+2\cot\theta)
 -\frac{1}{2}\frac{\sin^2\theta}{\sin^2 t}\ee^{2u}
 \left(Q_{,t}^2-Q_{,\theta}^2\right)=0.
\end{equation}
Alternatively to \eref{M}, the metric potential $M$ can also be
calculated from the first order field equations
\begin{eqnarray}\fl\label{M1a}
 (\cos^2t - \cos^2\theta)M_{,t}
 & = & \frac{1}{2}\ee^{2u}\frac{\sin^3\theta}{\sin t}\left[
       \cos t\sin\theta(Q_{,t}^2+Q_{,\theta}^2)
       -2\sin t\cos\theta\, Q_{,t}Q_{,\theta}\right]\nonumber\\
 && + \frac{1}{2}\sin t\sin\theta\left[\cos t\sin\theta(u_{,t}^2+u_{,\theta}^2)
       -2\sin t\cos\theta\, u_{,t}u_{,\theta}\right]\nonumber\\
 &&      +(2\cos^2t\cos^2\theta-\cos^2 t-\cos^2\theta)u_{,t}\nonumber\\
 &&   +2\sin t\cos t\sin\theta\cos\theta (u_{,\theta}-\tan\theta),       
\end{eqnarray}
\begin{eqnarray}\fl\label{M2a}
 (\cos^2t - \cos^2\theta)M_{,\theta}
 & = & -\frac{1}{2}\ee^{2u}\frac{\sin^3\theta}{\sin t}\left[
       \sin t\cos\theta(Q_{,t}^2+Q_{,\theta}^2)
       -2\cos t\sin\theta\, Q_{,t}Q_{,\theta}\right]\nonumber\\
 && - \frac{1}{2}\sin t\sin\theta\left[\sin t\cos\theta(u_{,t}^2+u_{,\theta}^2)
       -2\cos t\sin\theta\, u_{,t}u_{,\theta}\right]\nonumber\\
 &&    +2\sin t\cos t\sin\theta\cos\theta(u_{,t}+\tan t)\nonumber\\
 &&   +(2\cos^2t\cos^2\theta-\cos^2 t-\cos^2\theta)u_{,\theta}.       
\end{eqnarray}

These expressions tell us that (see \ref{App2} for a detailed derivation) 
\begin{equation}\label{BC1}
 M_{,t}=Q_{,t}=u_{,t}=0\,,\quad 
 Q = Q\p = \mbox{constant}\,,\quad  M+u=\textrm{constant}
\end{equation}
holds on $\Hp$. As the $t$-derivatives of all metric 
functions vanish identically at $\Hp$, a complete set of
initial data at $\Hp$ consists of
\begin{equation}\label{id}
  Q=Q\p\in\R,\quad u\in H^4,\quad Q_{,tt}\in H^2,
\end{equation}
where $Q_{,tt}$ is in $H^2$ as a consequence of the regularity assumptions discussed above.
Note that among the second $t$-derivatives only
$Q_{,tt}$ can be chosen freely since the values of $M_{,tt}$ as well
as $u_{,tt}$ are then fixed, as again the study of the field
equations \eref{E1}-\eref{M} near $\Hp$ reveals. 
Similarly, $M$ is also fixed on $\Hp$ by the choice of the data in
\eref{id}.

It turns out that the constant $\Qp$ is a gauge degree of freedom. This results from the fact that
the line element \eref{LE} is invariant under the coordinate change 
\beq\label{ccord_transf}
\Sigma:(t,\theta,\varphi,\delta) \mapsto \Sigma':(t,\theta,\varphi' =
\varphi -\Omega\delta,\delta), 
\eeq
leading to $Q\p'=Q\p+\Omega$ in the new coordinates \footnote{Note that 
for the corresponding black hole spacetimes, the coordinate change \eref{ccord_transf} describes a  
transformation into a rigidly rotating frame of reference (for more
details see  \cite{Ansorg2008,Ansorg2009,Hennig2009}).}.
We use this freedom in order 
to exclude two specific values, namely $Q\p=0$ and $Q\p=1/J$, where $J$ 
is the already mentioned conserved quantity that will be introduced in \eref{J}.
This exclusion becomes necessary since the analysis carried out below 
breaks down if $Q\p$ takes one of these values. 

We note further that as another consequence of our regularity requirements, the following axis condition 
holds at least in a neighborhood of the points $A$ and $B$
(cf.~Fig.~\ref{GS}): 
\begin{equation}\label{BC2}
 \mathcal A_{1/2}:\qquad M=u.
\end{equation}
Moreover, at these points $A,B$ we have (see \ref{App2})
\begin{equation}\label{BC3}
 M_A=M_B=u_A=u_B.
\end{equation}
Note that solutions which are also $C^2$-regular up to and including
$\Hf$ satisfy corresponding conditions at the points $C$ and $D$.

\subsection{The Ernst equation}

In order to introduce the Ernst formulation of the Einstein
equations, we define the complex Ernst potential
\begin{equation}\label{EP}
 \E(t,\theta)=f(t,\theta)+\ii b(t,\theta),
\end{equation}
where the real part $f$ is given by
\begin{equation}\label{Re}
 f:=-\xi_i\xi^i=-\ee^{-u}\sin^2\!t-Q^2\ee^{u}\sin^2\!\theta
\end{equation}
and the imaginary part $b$ is defined in terms of a potential $a$,
\begin{equation}\label{Im}
 a:=\frac{\xi^i\eta_i}{\xi^j\xi_j}=-\frac{Q}{f}\ee^u\sin^2\!\theta,
\end{equation}
via
\begin{equation}\label{a}
 a_{,t}      = \frac{1}{f^2}\sin t\sin\theta\, b_{,\theta},\qquad
 a_{,\theta} = \frac{1}{f^2}\sin t\sin\theta\, b_{,t}.
\end{equation}

In this formulation, the vacuum Einstein equations are equivalent to the
Ernst equation 
\begin{equation}\label{Ernst}
 \Re(\E)\left(-\E_{,tt}-\cot t\,\E_{,t}+\E_{,\theta\theta}
              +\cot\theta\,\E_{,\theta}\right)
  =-\E_{,t}^2+\E_{,\theta}^2,
\end{equation}
where $\Re(\E)$ denotes the real part of $\E$.
As a consequence of \eref{Ernst}, the integrability condition
$a_{,t\theta}=a_{,\theta t}$ of the system \eref{a} is satisfied such
that $a$ may be calculated from (\ref{a}) using $\E$. Moreover, given $a$ and $\E$ we
can use \eref{Re} and \eref{Im} to obtain the metric functions $u$ and
$Q$. Finally, the potential $M$ may be calculated
from
\begin{eqnarray}\label{M1}
 M_{,t} & = & -\frac{f_{,t}}{f} + \frac{1}{2f^2}
                   \frac{\sin t\sin\theta}{\cos^2\! t-\cos^2\!\theta}
                   \Big[\cos t\sin\theta
                         \left(f_{,t}^2+f_{,\theta}^2+b_{,t}^2
                       +b_{,\theta}^2\right)\nonumber\\
        & &                 -2\sin t\cos\theta
                         \left(f_{,t} f_{,\theta} + b_{,t} b_{,\theta}\right)
                         -4f^2\frac{\cos t}{\sin\theta}\Big],\\
 \label{M2}
 M_{,\theta} & = & -\frac{f_{,\theta}}{f} - \frac{1}{2f^2}
                   \frac{\sin t\sin\theta}{\cos^2\! t-\cos^2\!\theta}
                   \Big[\sin t\cos\theta
                         \left(f_{,t}^2+f_{,\theta}^2+b_{,t}^2
                               +b_{,\theta}^2\right)\nonumber\\
             & &            -2\cos t\sin\theta
                         \left(f_{,t} f_{,\theta} + b_{,t} b_{,\theta}\right)
                         -4f^2\frac{\cos\theta}{\sin t}\Big]
\end{eqnarray}
since the Ernst equation \eref{Ernst} also ensures the
integrability condition $M_{,t\theta}=M_{,\theta t}$.

As for the potentials introduced in Sec.~\ref{Subsec:coords} we
conclude axis conditions which 
hold at least in a neighborhood of the points $A$ and $B$ (cf.~Fig.~\ref{GS}):
\begin{equation}\label{BC2_E}
 \mathcal A_{1/2}:\qquad
 \E_{,\theta}=0,\quad a=0.
\end{equation}
Moreover, at the points $A,B$ we have $f=0$.
Again, solutions which are also $H^4$-regular on $\Hf$ satisfy
corresponding conditions at the points $C$ and $D$. 

It turns out that initial data
$\E\p(\theta)\equiv\E(0,\theta)=f\p(\theta)+\ii b\p(\theta)$ of the
Ernst potential are equivalent to the inital data set consisting of $u$, $Q=Q\p$, $Q_{,tt}$ at $\Hp$. 
Both sets are related via
\begin{eqnarray}\label{fp}
 f\p & = & -Q\p^2\ee^{u(0,\theta)}\sin^2\!\theta,\\
 b\p & = & b_A+2Q\p(\cos\theta-1)-Q\p^2\int_0^\theta
           \ee^{2u(0,\theta')}Q_{,tt}(0,\theta')\sin^3\!\theta'\,
           \dd\theta',
\end{eqnarray}
where $b_A=b(0,0)$ is an arbitrary integration constant.

\subsection{Conserved quantities\label{Sec:conserved}}

As a consequence of the symmetries of the Gowdy metric, there exist
{\em conserved} quantities, i.e.~integrals with respect to $\theta$ 
that are independent of the coordinate time $t$.
One of them is $J$, defined by
\begin{equation}\label{def_J}
 J:=-\frac{1}{8}\int_0^\pi\frac{Q_{,t}(t,\theta)}
 {\sin t}\,\ee^{2u(t,\theta)}\sin^3\!\theta\,\dd\theta = \mbox{constant}.
\end{equation}
As for the black hole angular momentum in the 
corresponding axisymmetric and stationary black hole 
spacetimes (cf.~discussion at the end of Sec.~\ref{sec:Intro}), 
this quantity determines whether or not a regular future Cauchy horizon exists.
In fact, it exists if and only if $J\neq 0$ holds. Note that $J$ vanishes in polarized
Gowdy models, where we have $Q_{,t}\equiv 0$.

It turns out that $J$ can be read off directly from the Ernst potential
and its second $\theta$-derivative at the points $A$ and $B$ on $\Hp$
(see Fig.~\ref{GS}), 
\begin{equation}\label{J}
  J = -\frac{1}{8Q\p^2}(b_A-b_B-4Q\p),\quad
  Q\p = -\frac{1}{2}b_{,\theta\theta}|_A
\end{equation}
where
\[
b_B = b(t=0,\theta=\pi).
\]

A detailed derivation of these formulas can be found in \cite{Hennig2009}.

\section{Potentials on $\Al$, $\Ar$, and $\Hf$\label{Sec:EP}}

\subsection{Ernst potential}

In the previous sections we have derived a formulation which permits
the direct translation to the situation in which the hyperbolic region 
inside the event horizon of an axisymmetric and stationary black hole 
(with possibly non-pure vacuum exterior, e.g.~with surrounding matter)
is considered, as was done in \cite{Ansorg2008,Ansorg2009,Hennig2009}.

In \cite{Ansorg2008} it has been demonstrated that a 
specific soliton method (the {\em B\"acklund transformation}, see \ref{App}) 
can be used to write the Ernst potential $\E$ in terms of another Ernst 
potential $\E_0$ which corresponds to a spacetime without a black hole, but
with a completely regular central vacuum region. Interestingly, the potential 
$\E_0=\E_0(t,\theta)$ possesses specific symmetry conditions which
translate here into 
\[
\begin{array}{lcll}
\E_0 (t, 0)      &=& \E_0 (0 , t)          &\quad \mbox{potential at $\Al$},\\[2mm]
\E_0 (t, \pi)    &=& \E_0 (0 , \pi-t)      &\quad \mbox{potential at $\Ar$},\\[2mm]
\E_0 (\pi , \theta) &=& \E_0 (0 , \pi-\theta) &\quad \mbox{potential at $\Hf$}.
\end{array}
\]
Hence the potential values at the boundaries $\Al$, $\Ar$ and $\Hf$ are
given explicitly in terms of those at $\Hp$. 
Now the B\"acklund transformation carries these dependencies over to the
corresponding original Ernst potential $\E$,  
i.e. we obtain $\E$ at $\Al$, $\Ar$ and $\Hf$ completely in terms of the
initial data at $\Hp$.

An alternative approach (see \cite{Ansorg2009,Hennig2009}) uses the \emph{inverse scattering method}. 
In these papers the
potentials on $\Al$, $\Ar$ and $\Hf$ were obtained from the
investigation of an associated linear matrix problem. The integrability conditions of 
this matrix problem are equivalent to the non-linear
field equations, see \ref{App}. 
We may carry the corresponding procedure over to our considerations of Gowdy spacetimes. 
Accordingly we are able to perform an explicit integration 
of the linear problem along the
boundaries of the Gowdy square. Since the resulting solution
is closely related to the Ernst potential,
it provides us with the desired expressions
between the metric quantities on the four boundaries of the Gowdy square. 

Note that in both approaches the axes $\Al$ and $\Ar$ are considered first. 
Starting at $\Hp$ and using the theorem by Chru\'sciel
\cite{Chrusciel1990}, which ensures $H^4$-regularity of the metric inside the
Gowdy square  
(i.e.~excluding only $\Hf$), we derive first the Ernst potentials at $\Al$ and $\Ar$ in terms of the values at $\Hp$. 
It turns out that for $J\neq 0$ these formulas can be extended continuously to the points $C$ and $D$ at which $\Al$ and $\Ar$ meet $\Hf$ (cf.~Fig.~\ref{GS}). Moreover, with the values at $C$ and $D$ it is possible to proceed to $\Hf$, and in this way we eventually find an Ernst potential which is continuous along the entire boundary of the Gowdy square. As the theorem by Chru\'sciel 
ensures unique solvability of the Einstein equations inside the Gowdy square, we conclude that the $H^4$-regularity of the Ernst potential 
holds up to and including $\Hf$ which therefore turns out to be an $H^4$-regular future Cauchy horizon.

The resulting expressions of the Ernst potentials at the boundaries
$\Al$, $\Ar$ and $\Hf$ read 
\begin{eqnarray}
 \fl\label{EA1}
 \Al: && \quad \E_1(x):=\E(t=\arccos x, \theta=0)\hspace{6.2mm}
                       = \frac{\ii[b_A-2\Qp(x-1)]\E\p(x)+b_A^2}
                        {\E\p(x)-\ii[b_A+2\Qp(x-1)]},\\
 \fl\label{EA2}
 \Ar: && \quad \E_2(x):=\E(t=\arccos(-x),\theta=\pi)
                       =  \frac{\ii[b_B-2\Qp(x+1)]\E\p(x)+b_B^2}
                        {\E\p(x)-\ii[b_B+2\Qp(x+1)]},\\
 \fl\label{EHf}
 \Hf: && \quad \E\f(x):=\E(t=\pi,\theta=\arccos(-x))\hspace{0.67mm}
                       = \frac{a_1(x)\E\p(x)+a_2(x)}{b_1(x)\E\p(x)+b_2(x)},  
\end{eqnarray}
where
\begin{equation}
 \E\p(x):=\E(t=0,\theta=\arccos x)
\end{equation} 
denotes the Ernst potential on
$\Hp$ and $a_1$, $a_2$, $b_1$, and $b_2$ in \eref{EHf} are
polynomials in $x$, defined by
\begin{eqnarray}\label{a1_b2}
 a_1 & = & \ii\big[16\Qp^2(1-x^2)+8\Qp(b_A(x+1)+b_B(x-1))\nonumber\\
     && \qquad +(b_A-b_B)(b_A(x-1)^2-b_B(x+1)^2)\big],\\
 a_2 & = & 8\Qp[b_A^2(x+1)+b_B^2(x-1)]-4b_Ab_B(b_A-b_B)x,\\
 \label{b1} 
 b_1 & = & 4(4\Qp+b_B-b_A)x,\\
 \label{b2} 
 b_2 & = & \ii\big[4\Qp(1-x^2)-b_A(1+x)^2+b_B(1-x)^2\big]
           (4\Qp+b_B-b_A).
\end{eqnarray}

A discussion of \eref{EHf} shows that $\E\f$ is indeed always
regular
provided that the black hole angular momentum does not vanish, which in turn means that 
$J\neq 0$, cf.~\eref{def_J}.
In order to prove this statement, we first note that both numerator and denominator on
the right hand side of \eref{EHf} are completely regular functions in terms of $x$, since
$a_1$, $a_2$, $b_1$, $b_2$ are polynomials in $x$ and the initial function $\E\p$
is regular by assumption. Hence, an irregular behavior of the potential $\E\f$ could 
only be caused by a zero of the denominator. Consequently, we investigate whether the equation
\begin{equation}\label{denom}
 b_1(x)\E\p(x)+b_2(x)=0
\end{equation}
has solutions $x\in[-1,1]$.
The real part of \eref{denom} is given by
\begin{equation}\label{denomr}
 4x(4Q\p+b_B-b_A)f\p(x)=0.
\end{equation}
Using \eref{fp} and \eref{J} together with our gauge $Q\p\neq0$ and the assumption $J\neq0$
we find that \eref{denomr} has exactly the three zeros, $x=-1$, $x=0$ and $x=1$
(corresponding to $\theta=\pi$,
$\theta=\pi/2$ and $\theta=0$). Now, for $x=0$ the imaginary part of
\eref{denom} does not vanish, whereas for $x=\pm 1$ it does.
Thus we find that the only zeros of the
denominator in \eref{EHf} are located at the two axes ($x=\pm 1$). 
As a matter of fact, the regular numerator of \eref{EHf} also vanishes 
at $x=\pm1$, as can be derived in a similar manner. Consequently, we study the behavior of 
$\E\f$ for $x=\pm 1$ in terms of the rule by L'H\^opital. As both numerator and denominator in \eref{EHf} have non-vanishing
values of the derivative with respect to $x$ for $x=\pm 1$, we conclude that the Ernst potential is regular everywhere
whenever $J\neq0$ holds.

Consider now the limit $J\to0$ for which the expression 
\[
4\Qp+b_B-b_A
\]
vanishes, cf.~\eref{J}. As this term appears as a factor in both $b_1$ and $b_2$ (cf. \eref{b1},\eref{b2}),
we find that the denominator in \eref{EHf} vanishes identically. 
The numerator, however, remains non-zero, which means that Ernst potential
diverges on the entire future boundary $t=\pi$, $0\le\theta\le\pi$.
We conclude that $\Hf$ becomes singular in the limit $J\to0$.
This divergent behavior of the Ernst potential corresponds to the
formation of a (scalar) curvature singularity at $\Hf$. In order to illustrate
this property, we calculate the Kretschmann scalar at the
point $C$ on $\Hf$ (see Fig.~\ref{GS}). 
Using the axis conditions discussed in Sec.~\ref{Sec:coords} and the
Einstein equations, we obtain 
\begin{equation}
 R_{ijkl}R^{ijkl}|_C
  =12\left[\ee^{-2u}(1+2u_{,tt})^2-Q_{,tt}^2\right]_C.
\end{equation} 
In terms of the Ernst potential, this expression reads
(cf. Eq.~\eref{FA1} below)
\begin{equation}
 R_{ijkl}R^{ijkl}|_C
 = \frac{1}{3}\left[(f_{,tt}+f_{,tttt})^2-(b_{,tt}+b_{,tttt})^2\right]_C.
\end{equation}
Now we can use \eref{EA1} to derive a formula that contains only the
initial data on the past horizon $\Hp$. Together with \eref{J} we get
\begin{equation}\label{Kret}\fl
 R_{ijkl}R^{ijkl}|_C = -\frac{3}{256Q\p^8
 J^6}\left[(16Q\p^4J^2-4b_{,xx}Q\p^2J-f_{,x}^2)^2
 -16Q\p^4J^2(f_{,xx}-2f_{,x})^2\right]_B,
\end{equation}
where $x=\cos\theta$.
Note that the numerator is well-defined and bounded for our
$H^4$-regular metric, a fact which is ensured by the validity of the
Einstein equations near $\Hp$. 

Equation \eref{Kret} indicates that the Kretschmann scalar diverges as
$J^{-6}$ in the limit $J\to0$. 
In fact, as we choose $Q\p\neq0$ (see Sec.~\ref{Subsec:coords}),
and furthermore $f_{,x}\neq0$ holds (because $2\pi f_{,x}=-Q\p^2A\p$
where $A\p$, $0<A\p<\infty$, is the horizon  area  
of $\Hp$, see Sec.~\ref{Sec:formula}), we conclude that 
$f_{,x}^4$ is the dominating term in the numerator of \eref{Kret} for
sufficiently small $J$.  
Hence the Kretschmann scalar indeed diverges as $J^{-6}$ in the limit $J\to0$. 

\subsection{Metric potentials}

From the Ernst potentials $\E_1=f_1+\ii b_1$, $\E_2=f_2+\ii b_2$,
$\E\f=f\f+\ii b\f$ in  \eref{EA1}, \eref{EA2}, \eref{EHf} we may
calculate the metric potentials $M$, $Q$ and $u$ on the boundaries of
the Gowdy square. Using \eref{Re}, \eref{Im}, \eref{a}, \eref{BC1},
\eref{BC2}, \eref{BC3} we obtain
\begin{eqnarray}
 \label{FA1}
 \Al: && \ee^{M_1}=\ee^{u_1}=-\frac{\sin^2\! t}{f_1},\quad
         Q_1 = \frac{b_{1,t}}{2\sin t},\\
 \Ar: && \ee^{M_2}= \ee^{u_2}=-\frac{\sin^2\! t}{f_2} ,\quad
         Q_2 = -\frac{b_{2,t}}{2\sin t},\quad\\
 \Hf: && \ee^{M\f}=-\frac{f_{,\theta\theta}^{\ 2}|_C}{4Q\f^2}
         \frac{\sin^2\!\theta}{f\f},\quad
         Q=Q\f,\quad
         \ee^{u\f} = -\frac{f\f}{Q\f^2\sin^2\!\theta},
\end{eqnarray}
where
\begin{eqnarray}
 Q_\mathrm{f} & = &
 \frac{b_A-b_B+4Q_\mathrm{p}}{b_A-b_B-4Q_\mathrm{p}}\,Q_\mathrm{p}.
\end{eqnarray}
Note that $Q_\mathrm{f} \neq 0$ in our gauge (cf.~\eref{J}):
\begin{equation*}\fl
 b_A-b_B+4Q_\mathrm{p}
 = (b_A-b_B-4Q_\mathrm{p})+8Q_\mathrm{p}=-8Q_\mathrm{p}^2J+Q_\mathrm{p}
 = 8Q_\mathrm{p}(1-JQ_\mathrm{p})\neq 0
\end{equation*}
in accordance with the discussion in Sec.~\ref{Subsec:coords} where the
gauge freedom was used to assure $0\neq Q_\mathrm{p}\neq 1/J$.
Furthermore, using \eref{EA1}-\eref{EHf} and our regularity assumptions for
the initial data, it is straightforward to show that $(-\sin^2\!t/f_1)$,
$(-\sin^2\!t/f_2)$ and $(-\sin^2\!\theta/f_f)$ are regular and positive
functions on the entire boundaries $\Al$, $\Ar$ and $\Hf$,
respectively. Moreover, the terms $(b_{1,t}/\sin t)$ and $(b_{2,t}/\sin t)$
are regular on $\Al$ and $\Ar$, respectively.
Consequently, the above boundary values for the
metric potentials $M$, $u$ and $Q$ are regular, too.

\section{A universal formula for the horizon areas\label{Sec:formula}}

In \cite{Ansorg2008} a relation between the black hole angular momentum 
and the two horizon areas of the outer event and inner Cauchy horizons
was found. This relation  
emerged from the explicit expressions of the inner Cauchy horizon potentials in terms of those 
at the event horizon. Translated to the case of general
$S^2\times S^1$ Gowdy spacetimes, 
this relation is given by 
\begin{equation}\label{area_formula}
 A\p A\f=(8\pi J)^2,
\end{equation}
where the areas $A_\mathrm{p}$
and $A_\mathrm{f}$ of the Cauchy horizons $\Hp$ and $\Hf$ are defined 
as integrals over the horizons (in a slice
$\delta=\textrm{constant}$),
\begin{equation}
 A_\mathrm{p/f} =
 \int\limits_{S^2}\sqrt{g_{\theta\theta}g_{\varphi\varphi}}\,\dd\theta\dd\varphi = 
 2\pi\int\limits_0^\pi\ee^{\frac{M+u}{2}}\big|_{\mathcal H_\mathrm{p/f}}
 \sin\theta\,\dd\theta=4\pi\ee^u|_{A/C}.
\end{equation}
	
\section{Discussion\label{Sec:Disc}}

In this paper we have analyzed general
$S^2\times S^1$ Gowdy models with a past 
Cauchy horizon $\Hp$. As any such spacetime can be related to a 
corresponding axisymmetric and stationary black hole solution,
considered between outer event and inner Cauchy horizons, the
results on the regularity of the interior of such black holes 
(obtained in \cite{Ansorg2008,Ansorg2009,Hennig2009}) can be carried over to the
Gowdy spacetimes treated here. In particular, specific soliton methods have proved to be useful, 
(i) the \emph{B\"acklund transformation} and (ii) the \emph{inverse scattering method}.
Both methods imply explicit expressions for the metric potentials on the boundaries
$\Al$, $\Ar$, $\Hf$ of the Gowdy square in terms of the initial values at $\Hp$. 
Moreover we obtain statements on existence and regularity of a future Cauchy
horizon as well as a universal relation for the horizon areas. These
results are summarized in the following.\\

{\noindent\bf Theorem 1.}
{\it Consider an $S^2\times S^1$ Gowdy spacetime with a past
Cauchy horizon~$\Hp$,
where the metric potentials $M$, $u$ and $Q$ appearing in the line element
\eref{LE2} are $H^4$-functions and the time derivatives
$H^3$-functions of the adapted coordinate $\theta$ on all slices
$t=\textrm{constant}$ in a closed neighborhood $N:=[0,t_0]\times[0,\pi]$,
$t_0\in(0,\pi)$, of $\Hp$. In addition, suppose $M,Q,u\in C^2(N)$.
Then this spacetime
possesses an $H^4$-regular future Cauchy horizon $\Hf$ if and only if
the conserved quantity $J$ (cf.~\eref{def_J}) does not vanish.
In the limit $J\to 0$, the future Cauchy horizon transforms into a
curvature singularity. Moreover, for $J\neq 0$ the universal relation
\begin{equation}\label{unirel}
 A_{\rm p} A_{\rm f} = (8\pi J)^2
\end{equation}
holds, where $A_{\rm p}$ and  $A_{\rm f}$ denote the areas of past and future
Cauchy horizons.}\\

\noindent\emph{Remark.} Note that the statements in Thm.~1 can be
generalized to $S^2\times S^1$ Gowdy spacetimes with additional electromagnetic
fields, see \cite{Ansorg2009,Hennig2009}. The proof
utilizes a more general linear matrix problem in which the Maxwell field 
is incorporated. Again the corresponding integrability conditions are equivalent 
to the coupled system of field equations that describe the Einstein-Maxwell field 
in electrovacuum with two Killing vectors (associated to the two Gowdy symmetries). 
It turns out that apart from $J$ a second conserved quantity $Q$ 
becomes relevant. The corresponding counterpart of this quantity 
in Einstein-Maxwell black hole spacetimes describes the electric charge of the black hole.
For Gowdy spacetimes we conclude that a regular future Cauchy horizon exists if and only
if $J$ and $Q$ do not vanish simultaneously. Moreover, we find that
Eq.~\eref{unirel} generalizes to $A\p A\f=(8\pi J)^2 +(4\pi Q^2)^2$.\\

With the above theorem we provide a long outstanding 
result on the existence of a regular future Cauchy horizon in $S^2\times S ^1$ Gowdy spacetimes. 
We note that the soliton methods being utilized in order to derive our conclusions
are not widely used in previous studies of this kind. Therefore we believe 
that these techniques might enhance further investigations in the 
realm of Gowdy cosmologies.

\ack
We would like to thank Florian Beyer and Piotr T. Chru\'sciel
for many valuable discussions and
John Head for commenting on the manuscript.
This work was supported by the Deutsche
For\-schungsgemeinschaft (DFG) through the
Collaborative Research Centre SFB/TR7
``Gravitational wave astronomy''.
\appendix
\section{Derivation of initial and boundary conditions\label{App2}}

We provide a derivation of the initial and boundary conditions through a thorough study of the Einstein equations
near the past Cauchy horizon~$\Hp$, i.e.~the initial surface $t=0$. 
First multiply the field equation \eref{E2} with $\sin t$ and
consider subsequently the limit $t\to0$. Taking our 
regularity assumptions into account 
(cf.~discusscion in Sec.~\ref{Subsec:coords}), we arrive at
\begin{equation}\label{AQ}
 Q_{,t}=0 \quad\textrm{at}\quad t=0.
\end{equation}
With the result \eref{AQ}, we study the limit $t\to 0$ of \eref{E2} in terms of the rule by L'H\^opital and  find
\begin{equation}
 \sin^3\theta\,\ee^{2u}Q_{,\theta}=\textrm{constant}.
\end{equation}
Evaluation at $\theta=0$ shows that the constant vanishes, leading to
\begin{equation}\label{AQ1}
 Q=\textrm{constant}\quad\textrm{at}\quad t=0.
\end{equation}
Next multiply Eq. \eref{E1} with $\sin t$ and study the limit $t\to 0$. 
With \eref{AQ} and \eref{AQ1} we obtain 
\begin{equation}
 u_{,t}=0\quad\textrm{for}\quad t=0.
\end{equation}
Using the previous results, we derive from \eref{M1a} in the 
limit $t\to0$
\begin{equation}
 M_{,t}=0\quad\textrm{at}\quad t=0.
\end{equation}
On the other hand, \eref{M2a} leads to
\begin{equation}\label{Anh3}
 M+u=\textrm{constant}\quad\textrm{at}\quad t=0.
\end{equation}

Similarly, we study the Einstein equations on the axis, i.e. in the limit
$\sin\theta\to 0$.
Multiplication of Eqs.~\eref{E1} and \eref{E2} with $\sin\theta$ leads
for $\sin\theta\to 0$ to
\begin{equation}\label{Anh1}
 u_{,\theta}=0,\quad Q_{,\theta}=0\quad\textrm{for}\quad \sin\theta=0.
\end{equation}
From \eref{M1a} and \eref{M2a} we obtain
\begin{equation}
\label{Mmu}
 M_{,\theta}=0,\quad M-u=\textrm{constant}\quad\textrm{for}\quad \sin\theta=0.
\end{equation}
As a consequence of an {\em axis regularity condition} 
that excludes the appearance of struts or knots along the axis (see
\cite{ExactSolutions} for details),
it turns out that the constant in \eref{Mmu} vanishes. Hence we get 
\begin{equation}\label{Anh2}
 M=u\quad\textrm{for}\quad \sin\theta=0.
\end{equation}

At the points $A$ and $B$ (see
Fig.~\ref{GS}), \eref{Anh2} means that $M_A=u_A$ and $M_B=u_B$, and with
$M_A+u_A=M_B+u_B$ (cf. \eref{Anh3}) we conclude
\begin{equation}
 M_A=u_A=M_B=u_B.
\end{equation}

Finally, we derive 
\begin{equation}
 \E_{,\theta}=0\quad\textrm{for}\quad\sin\theta=0
\end{equation}
by multiplying the Ernst equation \eref{Ernst} with $\sin\theta$ and considering the limit $\sin\theta\to0$.
Moreover, it follows from the definition \eref{Im} of the potential $a$
that
\begin{equation}
 a=0\quad\textrm{for}\quad\sin\theta=0.
\end{equation}

\section{The linear problem and B\"acklund transformations\label{App}}

In this appendix we briefly discuss the mathematical structure 
of the Ernst equation \eref{Ernst} which permits the application of so-called soliton methods. 
More details can be found in \cite{Ansorg2008,Ansorg2009,Hennig2009}. For a sophisticated introduction
to soliton methods for the axisymmetric and stationary Einstein equations we refer the reader to \cite{Neugebauer1996}.  

There are two soliton methods which lie at the heart 
of the treatment of $S^2\times S^1$ Gowdy spacetimes pursued in this paper:
(i) the \emph{B\"acklund transformation} and (ii)~the \emph{inverse scattering method}. Both methods 
make use of the following linear matrix problem (see \cite{Neugebauer1979,Neugebauer1980}), 
which read in our coordinates as follows:

\begin{equation}\label{LP}
\eqalign{
 \Phibold_{,x} & = \left[\left(\begin{array}{cc}
                   B_x & 0\\ 0 & A_x\end{array}\right)
                   +\lambda\left(\begin{array}{cc}
                   0 & B_x\\ A_x & 0\end{array}\right)\right]\Phibold,\\
 \Phibold_{,y} & = \left[\left(\begin{array}{cc}
                   B_y & 0\\ 0 & A_y\end{array}\right)
                   +\frac{1}{\lambda}\left(\begin{array}{cc}
                   0 & B_y\\ A_y & 0\end{array}\right)\right]\Phibold.        
}
\end{equation}
Here, $\Phibold=\Phibold(x,y,K)$ is a $2\times 2$ matrix pseudopotential
depending on the coordinates
\begin{equation}
 x = \cos(t-\theta),\qquad y = \cos(t+\theta)
\end{equation}
as well as on the \emph{spectral parameter} $K\in\C$. The function
$\lambda$ is defined as
\begin{equation}\label{lambda}
 \lambda(x,y,K) = \sqrt{\frac{K-y}{K-x}}.
\end{equation}
For fixed values $x$, $y$, the equation \eref{lambda} describes a mapping
$\C\to\C$, $K\mapsto\lambda$ from a two-sheeted Riemann surface
($K$-plane) onto the complex $\lambda$-plane.
In the $K$-plane the two $K$-sheets are connected at the branch points
\begin{equation}
 K_1 = x\quad (\lambda=\infty),\qquad K_2=y\quad (\lambda=0).
\end{equation}
Examining the integrability conditions $\Phibold_{,xy}=\Phibold_{,yx}$ yields, on the one hand, that the quantities
$A_x$, $A_y$, $B_x$ and $B_y$ are given in terms of a single complex
`Ernst' potential $\E=f+\ii b$,
\begin{equation}\label{A_i_B_i}
 A_i = \frac{\E_{,i}}{2f}, \qquad
 B_i = \frac{\bar\E_{,i}}{2f},\qquad i=x,y.
\end{equation}
On the other hand, the integrability conditions $\Phibold_{,xy}=\Phibold_{,yx}$ tell us 
that this potential $\E$ satisfies the Ernst equation \eref{Ernst}. 
Conversely, any solution $\E$ to the Ernst equation implies the existence
of an associated matrix $\Phibold$ which obeys the above linear matrix equations 
\eref{LP} where the functions $A_x$, $A_y$, $B_x$ and $B_y$ follow from \eref{A_i_B_i}.

Now, with a \emph{B\"acklund transformation} a new potential $\E$ can be constructed from a previously known one $\E_0$. 
Starting from $\E_0$ and the corresponding matrix function $\Phibold_0$, we consider transformations of the form
\begin{equation}\label{BT}
 \textrm{BT}_n:\quad \Phibold_0\mapsto \Phibold={\bf T}_n\Phibold_0,
 \quad n\in\N\ \textrm{even},
\end{equation}
where ${\bf T}_n$ is a matrix polynomial in $\lambda$ of degree $n$.
From $\Phibold$, determined via \eref{BT}, one can finally calculate the
corresponding new Ernst potential $\E$, see \cite{Neugebauer1996}.

Note that our specific expressions for the metric at the future Cauchy 
horizon $\Hf$ in Gowdy spacetimes can be obtained by considering the
particular case of a twofold B\"acklund transformation ($n=2$), for
which the new Ernst potential $\E$ reads 
\begin{equation}
 \E = \frac{\left[\alpha_1(\cos t+\cos\theta)-\alpha_2(\cos
       t-\cos\theta)\right]\E_0+2\bar\E_0}
       {\alpha_1(\cos t+\cos\theta)-\alpha_2(\cos
       t-\cos\theta)-2}.
\end{equation}
Here, $\alpha_1$ and $\alpha_2$ are solutions of the Riccati equations
\begin{eqnarray}
 &\alpha_{i,x}=&-(\lambda_i\alpha_i^2+\alpha_i)\frac{\E_{0,x}}{2f_0}
              +(\alpha_i+\lambda_i)\frac{\bar\E_{0,x}}{2f_0},\\
 &\alpha_{i,y}=&-\left(\frac{1}{\lambda_i}\alpha_i^2+\alpha_i\right)
               \frac{\E_{0,y}}{2f_0}
              +\left(\alpha_i+\frac{1}{\lambda_i}\right)
              \frac{\bar\E_{0,y}}{2f_0},\quad i=1,2,             
\end{eqnarray}
with
$$\alpha_i\bar\alpha_i=1,$$
where
$$\lambda_1:=\lambda(x,y,K=-1),\quad \lambda_2:=\lambda(x,y,K=1).$$

In our second approach, the \emph{inverse scattering method}, the linear problem 
\eref{LP} is integrated along the boundaries of the Gowdy square.
It turns out that explicit formulas can be found and that, moreover, 
the resulting solution must be continuous at 
this boundary (provided that the solution is regular at $\Hf$, which is
true for $J\neq0$, 
see discussion in Sec.~\ref{Sec:EP}). In this way we find the expressions that 
constitute the statements of this paper.

\section*{References}
%


\begin{thebibliography}{99}
\bibitem{Andersson}
Andersson L 2004
The global existence problem in general relativity
\emph{The Einstein equations and the large scale behavior of
gravitational fields: 50 years of the Cauchy problem in General 
Relativity} 
ed P T Chru\'sciel and H Friedrich (Basel, Boston: Birkh\"auser)
\bibitem{Ansorg2008}
Ansorg M and Hennig J 2008
\emph{Class. Quantum Grav.} {\bf 25} 222001 
\bibitem{Ansorg2009}
Ansorg M and Hennig J 2009
\emph{Phys. Rev. Lett.} {\bf 102} 221102
\bibitem{Bardeen}
Bardeen J M 1973  Rapidly rotating stars, disks, and black holes,
{\it Black holes (Les Houches)} ed C deWitt and B deWitt
(London: Gordon and Breach) pp 241-289
\bibitem{Berger1993}
Berger B K and Moncrief V 1993
\emph{Phys. Rev. D} {\bf 48} 4676
\bibitem{Beyer2008}
Beyer F 2008
\emph{Class. Quantum Grav.} {\bf 25} 235005
\bibitem{Beyer2009}
Beyer F 2009
\emph{J. Comput. Phys.} {\bf 228} 6496
\bibitem{Carter}
Carter B 1973   Black hole equilibrium states,
{\it Black holes (Les Houches)} ed C deWitt and B deWitt
(London: Gordon and Breach) pp 57 -- 214
\bibitem{Chandrasekhar}
Chandrasekhar S and Xanthopoulos B C 1986
\emph{Proc. R. Soc. A} {\bf 408} 175
\bibitem{Chrusciel1990}
Chru\'sciel P T 1990
\emph{Ann. Phys.} {\bf 202} 100
\bibitem{ChruscielIM}
Chru\'sciel P T, Isenberg J and Moncrief V 1990
\emph{Class. Quantum Grav.} {\bf 7} 1671
\bibitem{Chrusciel}
Chru\'sciel P T 2009 private communication
\bibitem{Clarke}
Clarke C J S 1993
\emph{The analysis of space-time singularities}
(Cambridge: Cambridge University Press)
\bibitem{Ellis}
Ellis G F R and Schmidt B G 1977
\emph{Gen. Relativ. Gravit.} {\bf 8} 915
\bibitem{Fischer}
Fischer A E 1970
The theory of superspace
\emph{Relativity ---
Proc. of the Relativity Conference in the Midwest}
ed M Carmeli, S I Fickler and L Witten
(New York: Plenum Press)
\bibitem{Garfinkle1999}
Garfinkle D 1999
\emph{Phys. Rev. D} {\bf 60} 104010
\bibitem{Gowdy1971}
Gowdy R H 1971
\emph{Phys. Rev. Lett.} {\bf 27} 826
\bibitem{Gowdy1974}
Gowdy R H 1974
\emph{Ann. Phys.} {\bf 83} 203
\bibitem{Griffiths}
Griffiths J B 1991
{\it Colliding plane waves in General Relativity}
(Oxford, New York, Tokyo: Clarendon Press)
\bibitem{Hawking}
Hawking S W and Ellis G F 1973
{\it The large scale structure of spacetime}
(Cambridge: Cambridge University Press)
\bibitem{Helliwell}
Helliwell T M and Konkowski D A 1999
\emph{Class Quantum Grav.} {\bf 16} 2709
\bibitem{Isenberg1990}
Isenberg J and Moncrief V 1990
\emph{Ann. Phys.} {\bf 199} 84
\bibitem{Hennig2009}
Hennig J and Ansorg M 2009
\emph{Ann. Henri Poincar\'e} {\bf 10} 1075
\bibitem{Kichenassamy1998}
Kichenassamy S and Rendall A D 1998
\emph{Class. Quantum Grav.} {\bf 15} 1339
\bibitem{Moncrief1981}
Moncrief V 1981
\emph{Ann. Phys.} {\bf 132} 87
\bibitem{Mostert}
Mostert P S 1957
\emph{Ann. Math.} {\bf 65} 447;
Erratum: {\bf 66} 589
\bibitem{Neugebauer1979}
Neugebauer G 1979
\emph{J.~Phys.~A} {\bf 12} L67
\bibitem{Neugebauer1980}
Neugebauer G 1980
\emph{J.~Phys.~A} {\bf 13} 1737
\bibitem{Neugebauer1996}
Neugebauer G 1996
Gravitostatics and rotating bodies
\emph{Proc. 46th Scottish Universities Summer School in Physics
(Aberdeen)}
ed G S Hall and J R Pulham
(London: Institute of Physics Publishing) pp 61-81
\bibitem{Neumann}
Neumann W D 1968
3-dimensional G-manifolds with 2-dimensional orbits
\emph{Proc. of the Conference on Transformation Groups}
ed P S Mostert, (Berlin, Heidelberg, New York: Springer) 
\bibitem{Obregon}
Obreg\'on O, Quevedo H and Ryan M P 2001
\emph{Phys. Rev. D} {\bf 65} 024022
\bibitem{Ringstrom2006}
Ringstr\"om H 2006
\emph{Comm. Pure Appl. Math.} {\bf 59} 977
\bibitem{Ringstrom2006b}
Ringstr\"om H 2009
\emph{Ann. Math.}
{\bf 170} 1181
\bibitem{Stahl}
St{\aa}hl F 2002
\emph{Class. Quantum Grav.} {\bf 19} 4483
\bibitem{ExactSolutions}
Stephani H, Kramer D, MacCallum M, Hoenselaers C, and Herlt E 2003
{\it Exact solutions of Einstein's field equations}
(Cambridge: Cambridge University Press).
\end{thebibliography}
\end{document}